\documentstyle[12pt]{article}
\begin{document}
\begin{flushright}
hep-th/0112260\\
SNBNCBS-2001
\end{flushright}
\vskip 3cm
\begin{center}
{\bf \Large { Topological aspects of Abelian gauge theory\\
in superfield formulation}}

\vskip 3cm

{\bf R.P.Malik}
\footnote{ E-mail address: malik@boson.bose.res.in  }\\
{\it S. N. Bose National Centre for Basic Sciences,} \\
{\it Block-JD, Sector-III, Salt Lake, Calcutta-700 098, India} \\

\vskip 3.5cm

\end{center}

\noindent
{\bf Abstract}:
We discuss some aspects of the topological features of a non-interacting
two $(1+1)$-dimensional Abelian gauge theory in the framework of superfield 
formalism. This theory is described by a BRST invariant Lagrangian density
in the Feynman gauge. We express the local and continuous symmetries,
Lagrangian density, topological invariants and symmetric energy momentum 
tensor of this theory in the language of superfields by exploiting the 
nilpotent (anti-)BRST- and (anti-)co-BRST symmetries. In particular, 
the Lagrangian density and symmetric energy momentum tensor 
of this topological theory turn out to be 
the sum of terms that geometrically correspond to the translations of some 
local superfields along the Grassmannian directions of the four 
($2+2$)-dimensional supermanifold. In this interpretation, the (anti-)BRST-
and (anti-)co-BRST symmetries, that emerge after the imposition of the
(dual) horizontality conditions, play a very important role.

\baselineskip=16pt


\newpage

\noindent
There are many areas of research in the modern developments of theoretical high
energy physics that have brought together mathematicians as well as theoretical
physicists to share their key insights into those specific fields of 
investigation
in a constructive and illuminating manner. The subject of topological field 
theories (TFTs) [1-3] is one such area that has provided a meeting-ground for 
both variety of researchers to enrich their understanding in a coherent and
consistent fashion. Recently, the free two ($1+1)$-dimensional (2D) Abelian- 
and self-interacting non-Abelian gauge theories (having no interaction with
matter fields) have been shown [4,5] to belong to a new class of TFTs that
capture together some of the key features of Witten- and Schwarz type
of TFTs [1,2]. Furthermore, these 2D free- as well as interacting 
(non-)Abelian gauge theories have been shown, in a series of papers [4-9], to
represent a class of field theoretical models for the Hodge theory where 
symmetries of the Lagrangian density and corresponding generators have been 
identified (algebraically) with the de Rham cohomology operators of
differential geometry. In fact, these symmetries and corresponding generators 
have been exploited to establish the topological nature of the 2D
free Abelian- and self-interacting non-Abelian gauge theories [5]. The
analogues of the above cohomological operators, in terms of the symmetries
and corresponding generators, have also been found for the physical four 
$(3 +1)$-dimensional free Abelian two-form gauge theory [10].
The geometrical interpretations for the above local and conserved
generators in the context of 2D theories
have also been provided [11-13] in the framework of the superfield 
formalism [14-18] where it has been shown that these conserved charges
correspond to the translation generators along the Grassmannian (odd)-
as well as bosonic (even) directions of a four ($2 + 2)$-dimensional
supermanifold. In these endeavours, a generalized version of the so-called
horizontality condition [14-16] has been exploited with respect to all 
the three 
\footnote{ On an ordinary flat manifold without a boundary, 
a set $(d, \delta, \Delta)$ of three cohomological operators can be
defined which obey the algebra:  $ d^2 = \delta^2 = 0, \Delta = (d + \delta)^2
= d \delta + \delta d \equiv \{ d, \delta \}, [\Delta, d] 
= [\Delta, \delta] = 0$ where $d = dx^\mu \partial_{\mu}$ and  $\delta 
= \pm * d *$ (with $*$ as the Hodge duality operation) are the nilpotent 
(of order two) exterior- and co-exterior derivatives and  $\Delta$ is the
Laplacian operator [19-22].} super de Rham cohomology operators 
$(\tilde d, \tilde \delta, \tilde \Delta= \tilde d
\tilde \delta + \tilde \delta \tilde d$) of differential geometry 
defined on the $(2 + 2)$-dimensional  
supermanifold without a boundary.

In all our previous attempts [11-13] to provide the geometrical interpretation
for the generators of the (anti-)BRST symmetries, (anti-)co-BRST symmetries and a bosonic symmetry in the framework of
superfield formulation, we have not found a way to capture the topological
features of the 2D free Abelian- and self-interacting non-Abelian gauge 
theories (without having any interaction with matter fields). The purpose
of our present paper is to show that the nilpotent ($s_{b}^2 = \bar s_{b}^2
= s_{d}^2 = \bar s_{d}^2 = 0)$ (anti-)BRST symmetries $(\bar s_{b})s_{b}$
and (anti-)co-BRST symmetries $(\bar s_{d})s_{d}$, Lagrangian density,
topological invariants and symmetric energy momentum tensor
for the free 2D Abelian gauge theory can be expressed in terms of the 
superfields alone and a possible geometrical interpretation can be provided for
the above physically
interesting  quantities in the framework of superfield formalism. We
show, in particular, that the Lagrangian density and the symmetric energy 
momentum tensor can be written as the sum of quantities that can be expressed
in terms of the Grassmannian derivatives on the 
Lorentz scalar(s) and second rank tensors, respectively. These scalar(s)
and tensors are constructed from the even superfields of the theory and
they are found to be endowed with the proper mass dimensions. In fact, 
for the present TFT (i.e. 2D free Abelian gauge theory),
the Lagrangian density and symmetric energy momentum tensor turn out to
have the geometrical interpretation as the sum of terms which correspond
to the translations of some local (but composite) even superfields 
(constructed by the basic even superfields of the theory) along the 
Grassmannian directions of the supermanifold. In a similar fashion,
the zero-forms of the topological invariants of this theory turn out to
be translations of the local (but composite) even superfields (constructed
by the basic odd superfields of the theory) along the Grassmannian directions 
of the $(2+2)$-dimensional supermanifold. These translations
are generated by the conserved and nilpotent (anti-) BRST- and (anti-)co-BRST 
charges.  One of the key features of this TFT
is the fact that the Lagrangian density and energy momentum tensor can be
expressed in terms of the even superfields  alone and the (anti-)BRST-
and (anti-)co-BRST transformations act on the $\theta\bar\theta$-components
of the one and the same combinations of the even superfields.
The symmetric
nature of the energy momentum tensor comes out very naturally in the framework
of superfield formulation. In the above derivations, the (dual) horizontality 
conditions w.r.t. super cohomological operators $\tilde d$ and $\tilde \delta$
play a very significant role. These conditions are, of course, required for
the derivations of the (anti-)BRST- and (anti-)co-BRST symmetries
which, in turn, provide the geometrical interpretation for their generators 
as the ``translation generators''  along the
Grassmannian ($\theta$ and $\bar \theta$) directions of the supermanifold. 
The superfield formulation of the above theory also sheds light on some
new symmetries of the Lagrangian density and symmetric energy momentum tensor
(see, e.g., equations (19b) and (38) below) which were not known hitherto
in our previous studies in the framework of Lagrangian formalism [4-10].

Let us begin with the BRST invariant Lagrangian density ${\cal L}_{b}$
for the free two ($1 + 1)$-dimensional
\footnote{We follow here the conventions and notations such that the 2D flat
Minkowski metric is: $\eta_{\mu\nu} =$ diag $(+1, -1)$ and $\Box = 
\eta^{\mu\nu} \partial_{\mu} \partial_{\nu} = (\partial_{0})^2 - 
(\partial_{1})^2, F_{01} 
= -\varepsilon^{\mu\nu} \partial_\mu A_\nu = E = \partial_{0} A_{1} 
- \partial_{1} A_{0}, \varepsilon_{01} =
\varepsilon^{10} = + 1, \; \varepsilon^{\mu\rho} \varepsilon_{\rho\nu}
= \delta_{\nu}^\mu.$ Here Greek 
indices: $\mu, \nu...= 0, 1$ correspond to the spacetime directions on
the 2D manifold.} 
Abelian gauge theory in the Feynman gauge [23-26]
$$
\begin{array}{lcl}
{\cal L}_{b} &=& - \frac{1}{4}\; F^{\mu\nu} F_{\mu\nu} 
- \frac{1}{2}\; (\partial \cdot A)^2 
- i \;\partial_{\mu} \bar C  \partial^\mu C \equiv
 \frac{1}{2}\; E^2
- \frac{1}{2} \; (\partial \cdot  A)^2
- i \;\partial_{\mu} \bar C \partial^\mu C
\end{array} \eqno(1)
$$
where $F_{\mu\nu} = \partial_\mu A_\nu - \partial_\nu A_\mu$ is the field
strength tensor derived from the connection one-form $ A = dx^\mu A_\mu$
(with $A_\mu$ as the vector potential) by application of the exterior
derivative $d = d x^\mu \partial_\mu $(with $d^2 = 0)$ as $ F = d A
= \frac{1}{2}\; (dx^\mu \wedge dx^\nu) F_{\mu\nu}$. The gauge-fixing term is 
derived as $\delta A = (\partial \cdot A)$ where $\delta = - * d * $ (with
$\delta^2 = 0$) is the co-exterior derivative and $*$ is the Hodge duality
operation. The (anti-)commuting ($C \bar C + \bar C C = 0, C^2 = \bar C^2 = 0)$
(anti-)ghost fields $(\bar C)C$ are required in the theory to maintain 
unitarity and gauge invariance together. The above Lagrangian density
(1) respects the following on-shell ($ \Box C = \Box \bar C = 0$)
nilpotent $(s_{b}^2 = 0,  s_{d}^2 = 0)$ BRST ($s_{b}$)
\footnote{We adopt here the notations and conventions of [26]. In fact,
in its full glory, a nilpotent ($\delta_{(D)B}^2 = 0$)
(co-)BRST transformation $(\delta_{(D)B})$ is equivalent 
to the product of an 
anti-commuting ($\eta C = - C \eta, \eta \bar C = - \bar C \eta$)
spacetime independent parameter $\eta$ and $(s_{d})s_{b}$ 
(i.e. $\delta_{(D)B} = \eta \; s_{(d)b}$) where $s_{(d)b}^2 = 0$.} 
-and dual(co)-BRST ($s_{d}$) symmetry transformations [4-9]
$$
\begin{array}{lcl}
s_{b} A_{\mu} &=& \partial_{\mu} C\; \qquad \;s_{b} C = 0 \; \qquad\; 
s_{b} \bar C = -i (\partial \cdot A)  \nonumber\\
s_{d} A_{\mu} &=& - \varepsilon_{\mu\nu} \partial^\nu \bar C\; \qquad\;
s_{d} \bar C = 0 \;\qquad \; s_{d} C = - i E.
\end{array}\eqno(2)
$$
The Lagrangian density (1) is also invariant under the on-shell anti-BRST- 
$(\bar s_{b})$ (with $s_{b} \bar s_{b} + \bar s_{b} s_{b} = 0$) and 
anti-co-BRST $(\bar s_{d})$ (with $s_{d} \bar s_{d} + \bar s_{d} s_{d}
= 0$) symmetries
$$
\begin{array}{lcl}
\bar s_{b} A_{\mu} &=& \partial_{\mu} \bar C\; \qquad \;\;\bar s_{b} \bar C 
= 0\; \qquad \;\;
\bar s_{b}  C = + \;i\; (\partial \cdot A)  \nonumber\\
\bar s_{d} A_{\mu} &=& - \varepsilon_{\mu\nu} \partial^\nu C\; \qquad
\bar s_{d}  C = 0\; \qquad \;\;\;\bar s_{d} \bar C = +\; i\; E.
\end{array} \eqno(3)
$$
The anti-commutator of these nilpotent, local, continuous and covariant
symmetries (i.e. $ s_{w} = \{s_{b}, s_{d}\}
= \{ \bar s_{b}, \bar s_{d}$) leads to a bosonic symmetry
\footnote{This symmetry has not been discussed in [27] where the
nilpotent transformations (2) and (3) have been discussed 
on a closed 2D Riemann surface. We thank Prof. N. Nakanishi for some
critical and constructive comments on our earlier works and for bringing
to our notice [27].}
$s_{w}$ ($ s_{w}^2 \neq 0$) transformations [4-9]
$$ 
\begin{array}{lcl}
s_{w} A_{\mu} &=&  \partial_{\mu} E
- \varepsilon_{\mu\nu} \partial^\nu (\partial \cdot A)\;
 \qquad s_{w} C = 0\; \qquad \;s_{w} \bar C = 0  
\end{array} \eqno(4)
$$
under which the Lagrangian density (1) transforms to a total derivative.
All the above continuous symmetry transformations can be concisely 
(and succinctly) expressed, in terms of the Noether
conserved charges $Q_{r}$ and $\bar Q_{r}$ [4-9], as
$$
\begin{array}{lcl}
s_{r} \Psi = - i \; [ \Psi, Q_{r} ]_{\pm} \quad 
Q_{r} = Q_{b}, Q_{d}, Q_{w}, Q_{g}, \quad
\bar s_{r} \Psi = - i \; [ \Psi, \bar Q_{r} ]_{\pm} \quad 
\bar Q_{r} = \bar Q_{b}, \bar Q_{d}
\end{array} \eqno(5)
$$
where brackets $[\;, \;]_{\pm}$ stand for the (anti-)commutators for 
any arbitrary generic field $\Psi$ being (fermionic)bosonic
in nature. Here the conserved ghost charge $Q_{g}$ generates
the continuous scale transformations: $ C \rightarrow e^{-\Sigma} C,
\bar C \rightarrow e^{\Sigma} \bar C, A_{\mu} \rightarrow A_{\mu},$ 
(where $\Sigma$ is a global parameter). The local 
field theoretical expressions for $Q_{r}$ and $\bar Q_{r}$
(which are not required for our present discussion) are given in [4-9].

The geometrical interpretation for the local and conserved
(anti-)BRST- $(\bar Q_{b})Q_{b}$ and (anti-)co-BRST $(\bar Q_{d})Q_{d}$
charges as the translation generators along the Grassmannian directions
of the ($2 + 2$)-dimensional supermanifold has
been shown [11-13] in the framework of superfield formulation [14-18] where
the even (bosonic) superfield $B_\mu (x,\theta,\bar\theta)$ and odd
(fermionic) fields $\Phi (x,\theta,\bar \theta)$ and $\bar \Phi(x,\theta,\bar
\theta)$ have been expanded in terms of the super coordinates
($x,\theta,\bar\theta)$, the dynamical fields of the Lagrangian density
(1) and some extra (secondary) fields (e.g.,$R_\mu (x), \bar R_\mu (x),
S_\mu (x), s(x), \bar s(x)$) as given below [11]
$$
\begin{array}{lcl}
B_{\mu} (x, \theta, \bar \theta) &=& A_{\mu} (x) 
+ \theta\; \bar R_{\mu} (x) + \bar \theta\; R_{\mu} (x) 
+ i \;\theta \;\bar \theta S_{\mu} (x) \nonumber\\
\Phi (x, \theta, \bar \theta) &=& C (x) 
+ i\; \theta \;(\partial \cdot A) (x)
- i \;\bar \theta\; E (x) 
+ i\; \theta\; \bar \theta \;s (x) \nonumber\\
\bar \Phi (x, \theta, \bar \theta) &=& \bar C (x) 
+ i \;\theta\;E (x) - i\; \bar \theta \;(\partial \cdot A) (x) 
+ i \;\theta \;\bar \theta \;\bar s (x).
\end{array} \eqno(6)
$$
Here some of the noteworthy points are: (i) the $(2+2)$-dimensional  
supermanifold is
parametrized by the superspace coordinates $Z^M = (x^\mu, \theta, \bar \theta)$
where $x^\mu (\mu = 0, 1)$ are the two even (bosonic) spacetime coordinates
and $\theta$ and $\bar \theta$ are the two odd (Grassmannian) coordinates
(with $\theta^2 = \bar \theta^2 = 0, 
\theta \bar \theta + \bar \theta \theta = 0)$. 
(ii) The expansions are along the odd (fermionic) superspace coordinates 
$\theta$ and $\bar \theta$ and even (bosonic)  $(\theta \bar\theta)$ 
directions of the supermanifold.  (iii) All the fields are local functions
of the spacetime coordinates $x^\mu$ alone (i.e.,$A_{\mu} (x,0,0) = A_\mu (x),
C (x,0,0) = C (x)$ etc.). Now the horizontality 
\footnote{ This condition is referred to as the ``soul-flatness'' condition
by Nakanishi and Ojima in [23].}
condition [14-16] on the super
curvature (two-form) tensor $\tilde F = \tilde d \tilde A$ for the Abelian
gauge theory
$$
\begin{array}{lcl} 
\tilde F =  \frac{1}{2}\; (d Z^M \wedge d Z^N)\;
\tilde F_{MN} = \tilde d \tilde A  \equiv 
d A = \frac{1}{2}\; (dx^\mu \wedge dx^\nu) \; F_{\mu\nu} = F
\end{array} \eqno(7)
$$
leads to the following expressions for the extra (secondary) fields [11]
$$
\begin{array}{lcl}
R_{\mu} \;(x) &=& \partial_{\mu}\; C(x)\; \qquad 
\bar R_{\mu}\; (x) = \partial_{\mu}\;
\bar C (x)\; \qquad\; \;s\; (x) = 0
\nonumber\\
S_{\mu}\; (x) &=& - \partial_{\mu} \;[(\partial \cdot A)] (x)
\qquad \;E \;(x) = 0\; \qquad \;\bar s \;(x) = 0
\end{array} \eqno(8)
$$
in terms of the basic fields (cf.Eqn.(1)) of the theory. The super curvature
tensor $\tilde F$ is constructed by the super exterior derivative 
$\tilde d$ and 
super connection one-form $\tilde A$, defined on the $(2+2)$-dimensional
supermanifold, as
$$
\begin{array}{lcl}
\tilde d &=& \;d Z^M \;\partial_{M} = d x^\mu\; \partial_\mu\;
+ \;d \theta \;\partial_{\theta}\; + \;d \bar \theta \;\partial_{\bar \theta}
\nonumber\\
\tilde A &=& d Z^M\; \tilde A_{M} = d x^\mu \;B_{\mu} (x , \theta, \bar \theta)
+ d \theta\; \bar \Phi (x, \theta, \bar \theta) + d \bar \theta\;
\Phi ( x, \theta, \bar \theta).
\end{array}\eqno(9)
$$
The substitution of (8) into expansion (6) leads to the following
$$
\begin{array}{lcl}
B_{\mu} (x, \theta, \bar \theta) &=& A_{\mu} (x) 
+ \theta\;  \partial_{\mu} \bar C (x) + \bar \theta\; \partial_{\mu} C (x) 
- i \;\theta \;\bar \theta \;\partial_{\mu} (\partial \cdot A) (x) \nonumber\\
&\equiv& A_\mu (x) + \theta\; (\bar s_{b} A_\mu (x)) + \bar \theta\;
(s_{b} A_\mu (x)) + \theta \;\bar \theta \;(s_{b} \bar s_{b} A_\mu (x))
\nonumber\\
\Phi (x, \theta, \bar \theta) &=& C (x) 
+ i\; \theta \;(\partial \cdot A) (x)
\equiv C (x) + \;\theta \;(\bar s_{b} C (x))  \nonumber\\
\bar \Phi (x, \theta, \bar \theta) &=& \bar C (x) 
-i\; \bar \theta \;(\partial \cdot A) (x)
\equiv \bar C (x)\; +\; \bar \theta \;(s_{b} \bar C (x)). 
\end{array} \eqno(10a)
$$
Thus, we notice that the horizontality condition in (7) leads to (i)
the derivation of secondary fields in terms of the basic fields of the
Lagrangian density. (ii) The (anti-)BRST symmetry transformations for the
Lagrangian density listed in (2) and (3). (iii) Geometrical interpretation
for the (anti-)BRST charges $(\bar Q_{b})Q_{b}$ as the translation generators
along the Grassmannian directions of the $(2+2)$-dimensional 
supermanifold, i.e.; 
$$
\begin{array}{lcl}
\mbox{Lim}_{\theta,\bar\theta \rightarrow 0}\; 
{\displaystyle \frac{\partial}{\partial \bar\theta}}\; B_\mu &=&
i\;[ Q_{b}, A_\mu ] \equiv s_{b} A_\mu \qquad
\mbox{Lim}_{\theta,\bar\theta \rightarrow 0}\; 
{\displaystyle \frac{\partial}{\partial \theta}}\; B_\mu =
i\;[\bar Q_{b}, A_\mu ] \equiv \bar s_{b} A_\mu \nonumber\\
\mbox{Lim}_{\theta,\bar\theta \rightarrow 0}\;
{\displaystyle \frac{\partial}{\partial \bar\theta}}\; \Phi &=&
- i\;\{ Q_{b}, C \} \equiv s_{b} C \qquad
\mbox{Lim}_{\theta,\bar\theta \rightarrow 0}\;
{\displaystyle \frac{\partial}{\partial \theta}}\; \Phi =
-i\;\{ \bar Q_{b}, C \} \equiv \bar s_{b} C \nonumber\\
\mbox{Lim}_{\theta,\bar\theta \rightarrow 0}\;
{\displaystyle \frac{\partial}{\partial \bar\theta}}\; \bar \Phi &=&
-i\;\{ Q_{b}, \bar C \} \equiv s_{b} \bar C  \qquad
\mbox{Lim}_{\theta,\bar\theta \rightarrow 0}\;
{\displaystyle \frac{\partial}{\partial \theta}}\; \bar \Phi =
-i\; \{ \bar Q_{b}, \bar C \} \equiv \bar s_{b} \bar C
\end{array}\eqno(10b)
$$
as is evident from equations (5) and (10a). It will
be noticed here that we have taken the translation generators
along the $\theta$- and $\bar \theta$ directions of the supermanifold 
as $ \frac{\partial}{\partial \theta}$ and
$ \frac{\partial} {\partial \bar \theta}$, respectively. (iv) The nilpotent 
(anti-)BRST transformations $(\bar s_{b})s_{b}$ are along the
Grassmannian directions $(\theta)\bar \theta$. (v) There is a mapping between
super exterior derivative $\tilde d$ and the (anti-)BRST charges as:
$\tilde d \Leftrightarrow (Q_{b}, \bar Q_{b})$. (vi) It is useful 
and interesting (for later convenience) to note that 
now the nilpotent (anti-)BRST symmetries of equations (2) and (3) can be 
re-written in terms of the superfields as 
$$
\begin{array}{lcl}
s_{b} B_{\mu} (x,\theta,\bar\theta)
&=& \partial_{\mu} \Phi (x,\theta,\bar\theta)
\; \;\;s_{b} \Phi (x,\theta,\bar\theta) = 0\; \;\;
s_{b} \bar \Phi (x,\theta,\bar\theta) = 
-i (\partial \cdot B) (x,\theta,\bar\theta)  \nonumber\\
\bar s_{b} B_{\mu} (x,\theta,\bar\theta)
&=& \partial_{\mu} \bar \Phi (x,\theta,\bar\theta)\; \;\; \bar s_{b} 
\bar \Phi (x,\theta,\bar\theta) = 0\; \;\; 
\bar s_{b} \bar \Phi (x,\theta, \bar\theta) 
= +\;i (\partial \cdot B) (x,\theta,\bar\theta)
\end{array}\eqno(11)
$$
where the expansions (10a) are taken into account which emerge after the
application of the horizontality condition w.r.t. the super exterior
derivative $\tilde d$. The sanctity and correctness of the above equation can 
be checked easily by first applying the transformations w.r.t. 
$\delta_{B} = \eta\; s_{b}$, 
and then, rederiving transformations $s_{b}$ from it.

The analogue 
\footnote{ Henceforth this condition w.r.t. (super) co-exterior derivatives
will be called as the dual horizontality condition because 
$(\tilde \delta) \delta$ and $(\tilde d) d$ are Hodge dual to each-other
on the (super) manifold.}
of the horizontality condition (7) w.r.t. the super co-exterior derivative 
$\tilde \delta = - \star \tilde d \star $ and its
operation on the super one-form connection $\tilde A$, namely;
$$
\begin{array}{lcl}
\tilde \delta \;\tilde A &=& \delta \; A \;\qquad\; \delta = - * d *\; \qquad\;
A = dx^\mu A_\mu \;\qquad \delta A = (\partial \cdot A) \nonumber\\
\tilde \delta \tilde A &=& (\partial_\mu B^\mu) + s^{\theta\theta} 
(\partial_{\theta} \Phi) + s^{\bar \theta \bar\theta} (\partial_{\bar\theta}
\bar \Phi) + s^{\theta \bar \theta} (\partial_{\theta} \bar \Phi +
\partial_{\bar \theta} \Phi)\nonumber\\
&-& \varepsilon^{\mu \theta} (\partial_{\mu} \Phi + \varepsilon_{\mu\nu} 
\partial_{\theta} B^\nu) - \varepsilon^{\mu \bar \theta} (\partial_{\mu}
\bar \Phi + \varepsilon_{\mu\nu} \partial_{\bar \theta} B^\nu)
\end{array} \eqno(12)
$$
leads to the following expression for the secondary (extra) fields in terms
of the basic fields of the Lagrangian density (1) for the theory [11,12]
$$
\begin{array}{lcl}
R_{\mu} (x) &=& - \varepsilon_{\mu\nu} \partial^\nu \bar C (x) 
\qquad
\bar R_{\mu} (x) = - \varepsilon_{\mu\nu} \partial^\nu  C (x)
\qquad s (x) = 0,   \nonumber\\
S_{\mu} (x) &=& + \varepsilon_{\mu\nu} \partial^\nu  E (x)\;
\;\;\qquad \;\;\;\bar s (x) = 0\; \;\;
\qquad \;\;\;(\partial \cdot A)\; (x) = 0.
\end{array} \eqno(13)
$$
In the above computations, the Hodge duality $\star$ operation
on the super differentials $(d Z^M)$ and their wedge products 
$(d Z^M \wedge d Z^N)$, defined on the $(2 + 2)$-dimensional supermanifold, 
is
$$
\begin{array}{lcl}
\star\; (dx^\mu) &=& \varepsilon^{\mu\nu}\; (d x_\nu)\; \qquad \;
\star\; (d\theta) = (d \bar \theta)\; \qquad \;
\star\; (d \bar \theta) = (d \theta) \nonumber\\
\star\; (d x^\mu \wedge d x^\nu) &=& \varepsilon^{\mu\nu}\; \qquad 
\star\; (dx^\mu \wedge d \theta) = \varepsilon^{\mu\theta}\; \qquad
\star\; (dx^\mu \wedge d \bar \theta) = \varepsilon^{\mu\bar\theta}
\nonumber\\
\star \; (d \theta \wedge d \theta) &=& s^{\theta\theta}\; \qquad
\star \; (d \theta \wedge d \bar \theta) = s^{\theta \bar \theta}\; \qquad
\star \; (d \bar \theta \wedge d \bar \theta) = s^{\bar \theta \bar \theta}
\end{array} \eqno(14)
$$
where $\varepsilon^{\mu\theta} = - \varepsilon^{\theta\mu}, \varepsilon^{\mu
\bar\theta} = - \varepsilon^{\bar\theta \mu},  
s^{\theta \bar \theta} = s^{\bar\theta\theta}$ etc. In terms of the expressions
(13), the super expansion (6) can be re-expressed as
$$
\begin{array}{lcl}
B_{\mu}\; (x, \theta, \bar \theta) &=& A_{\mu} (x) 
- \;\theta\; \varepsilon_{\mu\nu}\;\partial^{\nu}  C (x) 
- \;\bar \theta\;\varepsilon_{\mu\nu}\; \partial^{\nu} \bar C (x) 
+ \;i \;\theta \;\bar \theta \;
\varepsilon_{\mu\nu}\;\partial^{\nu}\; E (x) \nonumber\\
&\equiv& A_\mu (x) + \;\theta\; (\bar s_{d} A_\mu (x)) + \; \bar \theta\;
(s_{d} A_\mu (x)) + \;\theta\; \bar \theta\; (s_{d} \bar s_{d} A_\mu (x))
\nonumber\\
\Phi\; (x, \theta, \bar \theta) &=& C (x) 
- \;i \;\bar \theta \; E (x) \equiv C (x) + \;\bar \theta\; (s_{d} C(x))
\nonumber\\
\bar \Phi\; (x, \theta, \bar \theta) &=& \bar C (x) 
+\; i \; \theta\; E (x) \equiv \bar C (x) +\; \theta\; (\bar s_{d} \bar C(x)).
\end{array} \eqno(15)
$$
We pin-point some of the salient features of the nilpotent (anti-)co-BRST
symmetry transformations vis-a-vis 
(anti-)BRST symmetry transformations (and their generators).
The common features are:
(i) the (anti-)BRST- and (anti-)co-BRST symmetry
transformations are generated along the $\theta (\bar \theta)$ directions
of the supermanifold. (ii) Geometrically, 
the translation generators along the Grassmannian
directions of the supermanifold are the conserved and nilpotent
(anti-)BRST- and (anti-)co-BRST charges (cf.Eqn.(5)). (iii) 
For the odd (fermionic) superfields, the
translations are either along $\theta$ or $\bar \theta$ directions 
for the case of (anti-)BRST- and (anti-)co-BRST symmetries. 
(iv) For the  bosonic superfield, the translations are along both $\theta$ 
as well as $\bar \theta$ directions when we consider (anti-)BRST- and/or
(anti-)co-BRST symmetries. The key differences are:
(i) comparison between (10a) and (15) shows
that the (anti-)BRST transformations generate
translations along $(\theta)\bar \theta$
directions for the odd fields $(C)\bar C$.  In contrast, for the same fields,
 the (anti-)co-BRST transformations generate translations along
$(\bar\theta)\theta$ directions of the supermanifold. (ii) The restrictions 
$ \tilde \delta \tilde A = \delta A$ and $\tilde d \tilde A = d A$
(w.r.t. different cohomological operators)
produce (anti-)co-BRST- and (anti-)BRST symmetry transformations. 
(iii) The expressions
for $R_\mu$ and $\bar R_\mu$ in (8) and (13) are such that the kinetic
energy- and gauge-fixing terms of (1) remain invariant under (anti-)BRST-
and (anti-)co-BRST symmetries, respectively. (iv) It is very interesting
to note that the nilpotent (anti-)co-BRST transformations in (2) and (3) can 
now be re-expressed in terms of the superfields (analogous to equation (11)) as
$$
\begin{array}{lcl}
s_{d} B_{\mu} (x,\theta,\bar\theta)
&=& - \varepsilon_{\mu\nu} \partial^{\nu} \bar \Phi (x,\theta,\bar\theta)\; 
\qquad \;\;s_{d} \bar \Phi (x,\theta,\bar\theta) = 0 \nonumber\\
 s_{d}  \Phi (x,\theta,\bar\theta)
&=& + i \varepsilon^{\mu\nu} \partial_\mu 
B_\nu (x,\theta,\bar\theta) \qquad 
\; \bar s_{d}  \Phi (x,\theta,\bar\theta)= 0 \nonumber\\
\bar s_{d} B_{\mu} (x,\theta,\bar\theta)
&=& - \varepsilon_{\mu\nu} \partial^{\nu}  \Phi (x,\theta,\bar\theta)\;
\qquad\;\;
\bar s_{d}  \bar \Phi (x,\theta,\bar\theta)
= - i \varepsilon^{\mu\nu} \partial_\mu B_\nu (x,\theta,\bar\theta)  
\end{array}\eqno(16)
$$
where the expansions (15) have been taken into account that are obtained
after the imposition of the dual horizontality condition with
respect to the super co-exterior derivative $\tilde \delta$.
(v) For the (anti-)BRST- and (anti-)co-BRST symmetries
 the mapping are: $ \tilde d \Leftrightarrow (Q_{b}, \bar Q_{b}),
\tilde \delta \Leftrightarrow (Q_{d},\bar Q_{d})$ but the 
ordinary exterior- and co-exterior derivatives $d$ and $\delta$ are
 identified with $(Q_{b}, \bar Q_{d})$ and $(Q_{d}, \bar Q_{b})$
because of the ghost number considerations of a typical state 
in the quantum Hilbert space [4-7].

Exploiting equations (2), (3) and (5), it can be checked that the Lagrangian 
density (1) can be expressed, modulo some total derivatives, as
$$
\begin{array}{lcl}
{\cal L}_{b} &=& \{ Q_{d}, T_{1} \} + \{ Q_{b}, T_{2} \}
\equiv \{ \bar Q_{d}, P_{1} \} + \{\bar Q_{b}, P_{2} \}\;\nonumber\\
{\cal L}_{b} &=& s_{d} (i T_{1}) + s_{b} (i T_{2}) + \partial_{\mu} Y^\mu
\equiv \bar s_{d} (i P_{1}) + \bar s_{b} (i P_{2}) + \partial_{\mu} Y^\mu
\end{array} \eqno(17)
$$
where $T_{1} = \frac{1}{2} (EC), T_{2} = - \frac{1}{2} 
((\partial \cdot A)\bar C), P_{1} = - \frac{1}{2} (E\bar C),
P_{2} = \frac{1}{2} ((\partial \cdot A) C)$ and
$ Y^\mu = \frac{i}{2} (\partial^\mu \bar C C - \partial^\mu C \bar C)$. {\it 
The above Lagrangian density can also be understood as translations, generated
by the (anti-)BRST- and (anti-)co-BRST charges, along the Grassmannian
($\theta$ and $\bar \theta$) directions} of the supermanifold as given below
$$
\begin{array}{lcl}
{\cal L}_{b} &=& {\displaystyle \frac{i}{2}\;\frac{\partial}{\partial \theta}}
\;\Bigl [\;\{(\varepsilon^{\mu\nu} \partial_{\mu} 
B_\nu)\;\bar \Phi\}|_{\mbox{(anti-)co-BRST}} \;+ \; \{ 
(\partial\cdot B) \Phi \}|_{\mbox{(anti-)BRST}} \;\Bigr ]
\end{array}\eqno(18a)
$$
$$\begin{array}{lcl} 
{\cal L}_{b} &=& - {\displaystyle
\frac{i}{2}\;\frac{\partial}{\partial \bar \theta}}\;
\Bigl [ \; \{(\varepsilon^{\mu\nu} \partial_{\mu} 
B_\nu) \Phi \}|_{\mbox{(anti-)co-BRST}} \; + \; \{
(\partial\cdot B) \bar \Phi \}|_{\mbox{(anti-)BRST}}\;  \Bigr ] 
\end{array} \eqno(18b)
$$
where the subscripts (anti-)BRST- and (anti-)co-BRST stand for the insertion
of the expansions given in equations (10a) and (15), respectively. 
It is obvious that the expression for the Lagrangian density$
 {\cal L}_{b} = \{ \bar Q_{d}, P_{1} \} + \{\bar Q_{b}, P_{2} \}$ of equation
(17) is captured by (18a) and $
{\cal L}_{b} = \{ Q_{d}, T_{1} \} + \{ Q_{b}, T_{2} \}$ is captured by
(18b) in the language of the derivatives on the composite superfields
defined on the supermanifold. Geometrically, (18a) implies the translation
(by the translation generator $ \frac{\partial}{\partial \theta}$)
of the composite superfields $(\varepsilon^{\mu\nu} \partial_\mu B_\nu) \bar
\Phi$ and $(\partial \cdot B) \Phi$ along the $\theta$-direction of the
supermanifold. For this interpretation, the 
nilpotent (anti-)BRST- and (anti-)co-BRST 
symmetries, that emerge after the imposition of the (dual) 
horizontality conditions with respect to the super cohomological operator(s) 
$\tilde d$ (and $\tilde \delta$), play an important role. 
Similar interpretation can be associated with (18b) as well.
In terms of the superfield expansion in (6), we can re-express
the Lagrangian density (1) (or (17)) as
$$
\begin{array}{lcl}
{\cal L}_{b} &=& {\displaystyle \frac{i}{4} }\;
{\displaystyle \frac{\partial}{\partial \bar \theta}\;
\frac{\partial}{\partial \theta} \;\Bigl [ B_{\mu} (x,\theta,\bar\theta)
B^\mu (x,\theta,\bar\theta) \Bigr ]|_{\mbox{(anti-)BRST}}} \nonumber\\
&+& {\displaystyle \frac{i}{4}}\;
{\displaystyle \frac{\partial}{\partial \bar \theta}\;
\frac{\partial}{\partial \theta} \;\Bigl [ B_{\mu} (x,\theta,\bar\theta)
B^\mu (x,\theta,\bar\theta) \;\Bigr ]|_{\mbox{(anti-)co-BRST}}} \nonumber\\
&\equiv& - \frac{1}{2}\;\Bigl [i \bar R_\mu R^\mu + A^\mu S_{\mu}
\Bigr ]|_{\mbox{(anti-)BRST}}
 - \frac{1}{2}\;\Bigl [i \bar R_\mu R^\mu + A^\mu S_{\mu}
\Bigr ]|_{\mbox{(anti-)co-BRST,}}
\end{array} \eqno(19a)
$$
which turns out, in the language of symmetry
transformations, to be equivalent to
$$
\begin{array}{lcl}
{\cal L}_{b} = \frac{i}{4} \;s_{b} \;\bar s_{b}\;
\bigl (A_{\mu} (x) A^\mu (x) \bigr )
+ \frac{i}{4}\;s_{d}\;\bar s_{d}\;
\bigl( A_{\mu} (x) A^\mu (x) \bigr ).
\end{array} \eqno(19b)
$$
The subscripts (anti-)BRST- and (anti-)co-BRST 
in (19a) stand for the insertion
of the results from equations (8) and (13), respectively. In fact, the 
Lagrangian densities in (19a) 
and (19b) differ from the Lagrangian density (1) by a total
derivative: $ \frac{1}{2}\;\partial^{\mu} [ A_\mu (\partial
\cdot A) + \varepsilon_{\mu\nu} A^\nu E ]$. A few comments are in order.
First, it is evident that the $(\theta\bar\theta)$-component in the expansion
of the product $B_{\mu}(x,\theta,\bar\theta) B^\mu (x,\theta,\bar\theta)$
leads to the derivation of the Lagrangian density (1) as the sum of terms
on which the Grassmannian derivatives operate. Over and above, one has
to exploit the (anti-)BRST- and (anti-)co-BRST symmetries to obtain the
exact expression for the Lagrangian density (modulo some total derivatives).
Second, the horizontality
condition (7) and its analogue in (12) play a very important role in the
above derivation. Third, the geometrical interpretation for the Lagrangian
density (19a) can be thought of as being equivalent to a couple of successive
translations for the Lorentz super-scalar $B_\mu (x,\theta,\bar\theta)
B^\mu (x,\theta,\bar\theta)$ 
along the $\theta$- and $\bar\theta$ directions of the supermanifold.
Finally, {\it it appears to be an essential feature of a TFT
that the Lagrangian density can be expressed as the 
$\theta\bar\theta$-component of a Lorentz super-scalar that can be 
constructed by the even 
superfields of the theory. On this scalar, one has to apply (anti-)BRST-
and (anti-)co-BRST symmetries that emerge after the imposition of the (dual) 
horizontality conditions}.

Now let us concentrate on the topological invariants of the theory. For the
ordinary 2D manifold
\footnote{The 2D Minkowskian manifold is actually a non-compact manifold.
Thus, for the precise and accurate meaning of the topological invariants,
homology cycles, etc., one has to consider the Euclidean version of the
2D Minkowskian manifold which turns out to be a closed 2D Riemann surface.
Now the Greek indices $\mu,\nu,\rho...= 1, 2$ 
will imply the Euclidean directions and the flat metric on this manifold
will carry the same signs (unlike the opposite signs for the Minkowskian 
manifold). Such kind of analyses has been performed
in [27] for the 2D (non-)Abelian
gauge theories. For the sake of brevity, however,
we shall continue with the Minkowskian notations but shall keep in mind
this important fact and crucial point.}, 
there are four sets of such invariants w.r.t.
conserved ($\dot Q_{b} = \dot {\bar Q_{b}} = \dot Q_{d} 
= \dot {\bar Q_{d}} = 0$) and on-shell ($\Box C = \Box \bar C = 0$)
nilpotent $(Q_{b}^2 = \bar Q_{b}^2 = Q_{d}^2 = \bar Q_{d}^2 = 0)$ (anti-)BRST-
and (anti-) co-BRST charges. These are (for $ k = 0, 1, 2$)
$$
\begin{array}{lcl}
I_{k} = {\displaystyle \oint}_{C_{k}} V_{k} \qquad\;
\bar I_{k} = {\displaystyle \oint}_{C_{k}} \bar V_{k} \qquad\;
J_{k} = {\displaystyle \oint}_{C_{k}} W_{k} \qquad\;
\bar J_{k} = {\displaystyle \oint}_{C_{k}} \bar W_{k} 
\end{array} \eqno(20)
$$
where $C_{k}$ are the $k$-dimensional homology cycles in the 2D manifold and
$(\bar V_{k})V_{k}$ and $(\bar W_{k})W_{k}$ are the $k(= 0, 1, 2)$-forms
which are constructed w.r.t. (anti-)BRST charges $(\bar Q_{b})Q_{b}$ and
(anti-)co-BRST charges $(\bar Q_{d})Q_{d}$, respectively. The forms $V_{k}$
w.r.t. the nilpotent ($Q_{b}^2 = 0$) and
conserved ($\dot Q_{b} = 0$) BRST charge $Q_{b}$ are [5-7]
$$
\begin{array}{lcl}
V_{0} &=& - (\partial \cdot A)\; C\;  \qquad \;\;
V_{1} = \bigl [\; i C \partial_{\mu} \bar C - (\partial\cdot A) A_{\mu}
\;\bigr ]\; dx^\mu \nonumber\\
V_{2} &=& i \; \bigl [\; A_{\mu} \partial_{\nu} \bar C - \bar C
\partial_{\mu} A_{\nu} \; \bigr ]\; dx^\mu \wedge dx^\nu.
\end{array} \eqno(21)
$$
It is straightforward to check that forms $\bar V_{k}$ w.r.t. anti-BRST
charge $\bar Q_{b}$ can be obtained from the above by exploiting the discrete
symmetry transformations $ C \leftrightarrow \bar C,\;
(\partial \cdot A) \leftrightarrow - (\partial\cdot A)$
that connect BRST- and anti-BRST transformations in (2) and (3). The forms 
$W_{k}$ w.r.t. the co-BRST charge $Q_{d}$ are [5-7]
$$
\begin{array}{lcl}
W_{0} &=& E\; \bar C\;  \qquad \;\;
W_{1} = \bigl [\; \bar C \varepsilon_{\mu\rho}\partial^{\rho}  C 
- i E A_{\mu}
\;\bigr ]\; dx^\mu \nonumber\\
W_{2} &=& i \; \bigl [\; \varepsilon_{\mu\rho} \partial^{\rho}  C A_\nu\;
+ \;\frac{ C}{2} \varepsilon_{\mu\nu} (\partial \cdot A)
 \; \bigr ]\; dx^\mu \wedge dx^\nu
\end{array} \eqno(22)
$$
and $\bar W_{k}$ can be obtained from the above by the discrete symmetry
transformations: $ C \leftrightarrow \bar C,\;
E \leftrightarrow - E$ under which (anti-)co-BRST transformations in (2)
and (3) are connected with each-other. In the language of the superfields
$B_{\mu} (x,\theta,\bar\theta), \Phi (x,\theta,\bar \theta), \bar \Phi
(x,\theta,\bar\theta)$, the topological invariants in (21) can be recast
as the $\theta$ and $\bar \theta$ independent  components in
$$
\begin{array}{lcl}
V_{0} &=& - (\partial \cdot B)\; \Phi\;  \qquad \;\;
V_{1} = \bigl [\; i \Phi \partial_{\mu} \bar \Phi - (\partial\cdot B) B_{\mu}
\;\bigr ]\; dx^\mu \nonumber\\
V_{2} &=& i \; \bigl [\; B_{\mu} \partial_{\nu} \bar \Phi - \bar \Phi
\partial_{\mu} B_{\nu} \; \bigr ]\; dx^\mu \wedge dx^\nu
\end{array} \eqno(23)
$$
where we have to use the on-shell conditions $\Box \Phi = \Box \bar \Phi = 0,
\Box B_{\mu} = 0$ (which imply the validity of all the equations of motion
 $\Box C = \Box \bar C = \Box A_\mu = \Box (\partial \cdot A) =
\Box E = 0$ for the Lagrangian density (1)). Furthermore, we have to use the
expansions (10a)
which are obtained after the imposition of the horizontality condition (7).
In fact, we notice here that, to obtain the expressions for the topological 
invariants of the theory w.r.t (anti-)BRST charges $(\bar Q_{b})Q_{b}$
and (anti-)co-BRST charges $(\bar Q_{d})Q_{d}$
in terms of superfields, all one has to do is to replace: 
$$
\begin{array}{lcl} 
C \rightarrow \Phi \;\quad
\bar C \rightarrow \bar \Phi\;\;\; A_\mu \rightarrow B_\mu\; 
\quad (\partial\cdot A)
\rightarrow (\partial \cdot B)\;\;\;
 E = - \varepsilon^{\mu\nu} \partial_{\mu} 
A_{\nu} \rightarrow - \varepsilon^{\mu\nu} \partial_{\mu} B_{\nu}.
\end{array}\eqno(24)
$$ 
This straightforward substitution yields the desired
results here because the expansions in
(10a) and (15) (after the imposition of constraints $\tilde d \tilde A
= d A$ and $\tilde \delta \tilde A = \delta A$) are such that the analogue
of the transformations (2) and (3) are exactly imitated in terms of
superfields in equations (11) and (16), respectively. Even the on-shell
($\Box \Phi = \Box \bar \Phi = 0$) nilpotent properties of 
the (anti-)co-BRST- and (anti-)BRST
transformations in (16) and (11) are same as that of the ordinary ghost fields
(i.e., $\Box C = \Box \bar C = 0$). It is illuminating, however, to check
that the zero-forms $(\bar V_{0})V_{0}$ and $(\bar W_{0})W_{0}$ w.r.t.
(anti-)BRST- and (anti-)co-BRST charges can be computed directly from
the expansion of the  product of the superfields 
$\Phi (x,\theta,\bar\theta) \bar \Phi (x,\theta,\bar\theta)$
along the $\theta, \bar \theta$ and $\theta\bar\theta$ directions, namely;
$$
\begin{array}{lcl}
(\Phi \bar \Phi)|_{\mbox{(anti-)BRST}}
&=& C \bar C + i \;\theta\; \bar C \;(\partial \cdot A) + i\; \bar \theta\;
C\; (\partial\cdot A) + \theta\bar\theta\; (\partial\cdot A)^2 \nonumber\\
(\Phi \bar \Phi)|_{\mbox{(anti-)co-BRST}}
&=& C \bar C - i\; \theta\; C E - i \;\bar \theta\;
\bar C E  - \theta\bar\theta \;E^2
\end{array} \eqno(25)
$$
where the subscripts stand for the expansions in (10a) and (15) that are 
obtained after the imposition of the horizontality- and the analogue of
horizontality conditions in (7) and (12), respectively. Now, it is 
straightforward to check that
$$
\begin{array}{lcl}
i\; {\displaystyle \frac{\partial 
(\Phi\bar\Phi)|_{\mbox{(anti-)BRST}}}
{\partial \theta}} &=& \bar V_{0}\; \qquad\;
i\; {\displaystyle \frac{\partial (\Phi
\bar\Phi)|_{\mbox{(anti-)BRST}}}
{\partial \bar \theta}} =  V_{0} \nonumber\\
i\; {\displaystyle \frac{\partial (\Phi\bar\Phi)|_{\mbox{(anti-)co-BRST}}}
{\partial \theta}} &=& \bar W_{0}\; \qquad
i\; {\displaystyle \frac{\partial (\Phi\bar\Phi)|_{\mbox{(anti-)co-BRST}}}
{\partial \bar \theta}} =  W_{0}
\end{array} \eqno(26)
$$
leads to the zero-forms of equations (21) and (22). Thus, the zero-forms
in the expression for topological invariants find a geometrical
interpretation as the translations for the local (but composite)
superfields $(\Phi \bar \Phi) (x,\theta,\bar\theta)$ along the Grassmannian
directions ($\theta$- and $\bar\theta$) of the supermanifold. By construction,
these quantities are (anti-)BRST- and (anti-)co-BRST invariant.
From these expressions,
one can always compute rest of the topological invariants by exploiting
the following recursion relations [5]
$$
\begin{array}{lcl}
s_{b}\; V_{k} &=& d\; V_{k-1}\;
\qquad \bar s_{b}\; \bar V_{k} = d \;\bar V_{k-1}\;
\qquad d = dx^\mu \;\partial_{\mu} \nonumber\\
s_{b}\; W_{k} &=& \delta\; W_{k-1}\; \qquad \bar s_{b}\; \bar W_{k} 
= \delta \;\bar W_{k-1}\;
\quad \delta = i\;dx^\mu \varepsilon_{\mu\nu} \partial_{\nu}
\end{array} \eqno(27)
$$
where $k = 1, 2$. The above relations are one of the key features for
the existence of a TFT.

One of the central properties of a TFT is the lack of energy excitations
in the physical sector of the theory. This happens because of the fact
that when operator form of the Hamiltonian density ($T^{(00)}$) is
sandwiched between two physical states, it yields zero (see, e.g.,
[3]). Thus, the form
of the symmetric energy momentum tensor ($T^{(s)}_{\mu\nu}$) plays a
very important role in this discussion. For the Lagrangian density
(${\cal L}_{b}$) of equation (1), the explicit form of the this
symmetric tensor is [5-7]
$$
\begin{array}{lcl}
T^{(s)}_{\mu\nu} &=& - \frac{1}{2}\; (\partial \cdot A)\;
(\partial_{\mu} A_{\nu} + \partial_{\nu} A_{\mu} ) - \frac{1}{2}
\; E\; (\varepsilon_{\mu\rho} \partial_{\nu} A^\rho
 + \varepsilon_{\nu\rho} \partial_{\mu} A^\rho) \nonumber\\
&-& i\; \partial_{\mu} \bar C \partial_{\nu} C 
- i\; \partial_{\nu} \bar C \partial_{\mu} C  - \eta_{\mu\nu} {\cal L}_{b}.
\end{array} \eqno(28)
$$
With the help of (17) (together with transformations (2) and (3) and equation
(5)), it can be checked that the above equations can be written, modulo some 
total derivatives, as
$$
\begin{array}{lcl}
T^{(s)}_{\mu\nu} = \{ Q_{b}, S_{\mu\nu}^{(1)} \} 
+ \{ Q_{d}, S_{\mu\nu}^{(2)} \} \equiv
\{ \bar Q_{b}, \bar S_{\mu\nu}^{(1)} \} 
+ \{ \bar Q_{d}, \bar S_{\mu\nu}^{(2)} \}
\end{array} \eqno(29)
$$
where the local expressions for $S_{\mu\nu}^{(1,2)}$ and 
$\bar S_{\mu\nu}^{(1,2)}$ are
$$
\begin{array}{lcl}
S_{\mu\nu}^{(1)} &=& \frac{1}{2} \; \bigl [\;
(\partial_{\mu} \bar C) A_{\nu} 
+ (\partial_{\nu} \bar C) A_{\mu} + \eta_{\mu\nu} (\partial \cdot A)\; \bar C
\;\bigr ] \nonumber\\
S_{\mu\nu}^{(2)} &=& \frac{1}{2} \; \bigl [\;
(\partial_{\mu}  C) \varepsilon_{\nu\rho} A^{\rho} 
+ (\partial_{\nu}  C) \varepsilon_{\mu\rho} A^{\rho} 
- \eta_{\mu\nu} E  C \;\bigr ] \nonumber\\
\bar S_{\mu\nu}^{(1)} &=& - \frac{1}{2} \; \bigl [\;
(\partial_{\mu}  C) A_{\nu} 
+ (\partial_{\nu}  C) A_{\mu} + \eta_{\mu\nu} (\partial \cdot A)\;  C
\;\bigr ] \nonumber\\
\bar S_{\mu\nu}^{(2)} &=& - \frac{1}{2} \; \bigl [\;
(\partial_{\mu} \bar C) \varepsilon_{\nu\rho} A^{\rho} 
+ (\partial_{\nu}  \bar C) \varepsilon_{\mu\rho} A^{\rho} 
- \eta_{\mu\nu} E \bar C \;\bigr ].
\end{array} \eqno(30)
$$
We can exploit now the
finer details of the superfield expansions in (10a) and (15) to 
express the above $S's$ in terms of the superfields. Towards this goal,
it is first essential to express $T's$ and $P's$ of (17) in the
language of the superfields. It is straightforward to check, from the
product of the odd superfields in (25), that
$$
\begin{array}{lcl}
{\displaystyle \frac{i}{2}\; \frac{\partial}{\partial \theta}}
\;\bigl [\;\Phi (x,\theta,\bar\theta) \bar \Phi (x,\theta,\bar\theta) 
\;\bigr ]|_{\mbox{(anti-)BRST}} &=& - \frac{1}{2} (\partial \cdot A) 
\bar C = T_{2} \nonumber\\
- {\displaystyle \frac{i}{2}\; \frac{\partial}{\partial \bar \theta}}
\;\bigl [\;\Phi (x,\theta,\bar\theta) \bar \Phi (x,\theta,\bar\theta)
\;\bigr ]|_{\mbox{(anti-)BRST}} &=& + \frac{1}{2} (\partial \cdot A) 
 C = P_{2} \nonumber\\
{\displaystyle \frac{i}{2}\; \frac{\partial}{\partial \theta}}
\;\bigl [\; \Phi (x,\theta,\bar\theta) \bar \Phi (x,\theta,\bar\theta)
\;\bigr ]|_{\mbox{(anti-)co-BRST}} &=&  + \frac{1}{2} \;(E C) = T_{1}
\nonumber\\
- {\displaystyle \frac{i}{2}\; \frac{\partial}{\partial \bar \theta}}
\;\bigl [\; \Phi (x,\theta,\bar\theta) \bar \Phi (x,\theta,\bar\theta)
\;\bigr ]|_{\mbox{(anti-)co-BRST}} &=& - \frac{1}{2}\; (E \bar  C) = P_{1}
\end{array} \eqno(31)
$$
where the subscripts have the same interpretations as explained earlier
(after equation (25)). It will be noticed that these $T's$ and $P's$ are the
same (modulo some constant factors)
as the zero-forms (26) in the topological invariants. Thus, these
quantities have the same geometrical interpretation as the zero-forms of
the topological invariants.
Rest of the terms in $S_{\mu\nu}^{(1,2)}$ can be
written, in terms of superfields, as
$$
\begin{array}{lcl}
&&{\displaystyle \frac{1}{2}\; \frac{\partial}{\partial \theta}}
\;\bigl [\; B_{\mu} (x,\theta,\bar\theta) B_\nu (x,\theta,\bar\theta) 
\;\bigr ]|_{\mbox{(anti-)BRST}} =
\frac{1}{2}\; \bigl (A_\mu \bar R_\nu + \bar R_\mu A_\nu 
\bigr )|_{\mbox{(anti-)BRST}} \nonumber\\
&& - \frac{1}{2}\;\varepsilon_{\mu\rho}\; \varepsilon_{\nu\sigma}\;
{\displaystyle  \frac{\partial}{\partial \theta}}
\;\bigl [\; B^{\rho} (x,\theta,\bar\theta) B^\sigma (x,\theta,\bar\theta) 
\;\bigr ]|_{\mbox{(anti-)co-BRST}} \nonumber\\
&& = \;- \;\frac{1}{2}\; \varepsilon_{\mu\rho}\; \varepsilon_{\nu\sigma}
\bigl (A^\rho \bar R^\sigma + \bar R^\rho A^\sigma 
\bigr )|_{\mbox{(anti-)co-BRST.}}
\end{array} \eqno(32)
$$
The r.h.s. of the above equations can be expressed in terms of the 
gauge field $A_\mu$ and the (anti-)ghost fields $(\bar C)C$ as
$$
\begin{array}{lcl}
\frac{1}{2} \; \bigl [\;
(\partial_{\mu} \bar C) A_{\nu} 
+ (\partial_{\nu} \bar C) A_{\mu} 
\;\bigr ] \qquad \mbox{and} \qquad
\frac{1}{2} \; \bigl [\;
(\partial_{\mu}  C) \varepsilon_{\nu\rho} A^{\rho} 
+ (\partial_{\nu}  C) \varepsilon_{\mu\rho} A^{\rho} 
\; \bigr ]
\end{array} \eqno(33)
$$
respectively. Here in equation (33), we have substituted the values of 
$\bar R's$ from (8) and (13). This equation yields, modulo some 
total derivatives, the desired result. Ultimately, the expression for 
the $S_{\mu\nu}^{(1,2)}$ in terms of the superfields, are
$$
\begin{array}{lcl}
S_{\mu\nu}^{(1)} &=& {\displaystyle \frac{1}{2} \; \frac{\partial}
{\partial \theta}} \;\bigl [\; B_\mu (x,\theta,\bar \theta)
B_{\nu} (x,\theta,\bar\theta)\;
\bigr ]|_{\mbox{(anti-)BRST}} \nonumber\\
&-&  {\displaystyle \frac{i}{2}\; \eta_{\mu\nu}\;
\frac{\partial}{\partial \theta}}\; \bigl [ \;\Phi (x,\theta,\bar\theta)
\bar\Phi (x,\theta,\bar\theta)
\;\bigr ]|_{\mbox{(anti-)BRST}} \nonumber\\
S_{\mu\nu}^{(2)} &=& - \frac{1}{2}\;
\varepsilon_{\mu\rho}\; \varepsilon_{\nu\sigma}\;
{\displaystyle  \frac{\partial}
{\partial \theta}}\; \bigl [\; B^\rho (x,\theta,\bar \theta)
B^{\sigma} (x,\theta,\bar\theta)\;
\bigr ]|_{\mbox{(anti-)co-BRST}} \nonumber\\
&-& {\displaystyle \frac{i}{2}}\; \eta_{\mu\nu}\;
{\displaystyle \frac{\partial}{\partial \theta}} \;\bigl [\; \Phi (x,\theta,
\bar\theta) \bar\Phi (x,\theta,\bar\theta)
\;\bigr ]|_{\mbox{(anti-)co-BRST.}} 
\end{array} \eqno(34)
$$
Geometrically, the expression for $S_{\mu\nu}^{(1)}$ correspond to the
translation of a second-rank tensor $B_{\mu} (x,\theta,\bar\theta)
B_\nu (x,\theta,\bar\theta)$ (constructed by the even superfields)
plus another second-rank tensor $\eta_{\mu\nu} \Phi (x,\theta,\bar\theta)
\bar \Phi (x,\theta,\bar\theta)$ (constructed by the odd superfields)
along the $\theta$-direction of the supermanifold. Similar
interpretation can be attached to the local 
expression for $S_{\mu\nu}^{(2)}$.
The local expressions for $\bar S_{\mu\nu}^{(1,2)}$ can also be 
computed in terms of the superfields. In fact, these depend on the
derivative w.r.t. $\bar \theta$, as given below
$$
\begin{array}{lcl}
\bar S_{\mu\nu}^{(1)} &=& - {\displaystyle \frac{1}{2} \; \frac{\partial}
{\partial \bar \theta}}\;\; \bigl [\; B_\mu (x,\theta,\bar \theta)
B_{\nu} (x,\theta,\bar\theta)\;
\bigr ]|_{\mbox{(anti-)BRST}} \nonumber\\
&+& {\displaystyle  \frac{i}{2}\; \eta_{\mu\nu}\;
\frac{\partial}{\partial \bar \theta}}\; \;\bigl [\; \Phi (x,\theta,\bar\theta)
\bar\Phi (x,\theta,\bar\theta)
\;\bigr ]|_{\mbox{(anti-)BRST}} \nonumber\\
\bar S_{\mu\nu}^{(2)} &=& + {\displaystyle \frac{1}{2} \;
\varepsilon_{\mu\rho}\;\varepsilon_{\nu\sigma}\; \frac{\partial}
{\partial \bar \theta}} \;\;\bigl [\; B^\rho (x,\theta,\bar \theta)
B^\sigma (x,\theta,\bar\theta)\;
\bigr ]|_{\mbox{(anti-)co-BRST}} \nonumber\\
&+& {\displaystyle \frac{i}{2}\; \eta_{\mu\nu}
\frac{\partial}{\partial \bar \theta}} \;\bigl [\;\Phi (x,\theta,\bar\theta)
\bar\Phi (x,\theta,\bar\theta)
\;\bigr ]|_{\mbox{(anti-)co-BRST.}} 
\end{array} \eqno(35)
$$
The geometrical interpretation in the language of the ``translations''
can be given to the above expressions in the same way as that of their
counterparts in (34).
Finally, the expression for the symmetric energy momentum tensor in (28)
can be expressed in terms of the even superfields 
{\it alone} and the Grassmannian
derivatives on them, as 
$$
\begin{array}{lcl}
T_{\mu\nu}^{(s)} &=& 
{\displaystyle \frac{i}{2} \; \frac{\partial}{\partial \bar\theta}\;
\frac{\partial} {\partial \theta}} \;\bigl [\; B_\mu (x,\theta,\bar \theta)
B_{\nu} (x,\theta,\bar\theta)\;
\bigr ]|_{\mbox{(anti-)BRST}} \nonumber\\
&-& {\displaystyle \frac{i}{2} \; \varepsilon_{\mu\rho}\; 
\varepsilon_{\nu\sigma}\; \frac{\partial}{\partial \bar\theta}
\frac{\partial} {\partial \theta}} \;\bigl [\; B^\rho (x,\theta,\bar \theta)
B^\sigma (x,\theta,\bar\theta)\;
\bigr ]|_{\mbox{(anti-)co-BRST}} \nonumber\\
&-& {\displaystyle \frac{i}{4}\; \eta_{\mu\nu}\; 
 \frac{\partial}{\partial \bar \theta}\;
\frac{\partial}{\partial \theta}} \;\bigl [\; B_{\rho} (x,\theta,\bar\theta)
B^{\rho} (x,\theta,\bar\theta)
\;\bigr ]|_{\mbox{(anti-)BRST}} \nonumber\\
&-&  {\displaystyle \frac{i}{4}\;\eta_{\mu\nu}\;
\frac{\partial}{\partial \bar \theta}\;
\frac{\partial}{\partial \theta}}\; \bigl [\;B_{\rho} (x,\theta,\bar\theta)
B^{\rho} (x,\theta,\bar\theta)\;
\bigr ]|_{\mbox{(anti-)co-BRST}}
\end{array} \eqno(36)
$$
where the general expression for the first term in the above equation is
$$
\begin{array}{lcl}
{\displaystyle \frac{i}{2} \; \frac{\partial}{\partial \bar\theta}\;
\frac{\partial} {\partial \theta}} \;\bigl [\; B_\mu (x,\theta,\bar \theta)
B_{\nu} (x,\theta,\bar\theta)\;
\bigr ] = - \frac{1}{2}\; (A_\mu S_{\nu} + S_\mu A_\nu)
+ \frac{i}{2}\; (R_\mu \bar R_\nu - \bar R_{\mu} R_\nu).
\end{array} \eqno(37)
$$
In this derivation, the general form of the superfield expansion (6) has
been used. To obtain the exact form of the expression (28) for the symmetric
energy momentum tensor, one has to substitute in (37) the values of the
extra secondary fields $R_\mu, \bar R_\mu, S_\mu$ as quoted in equations
(8) and (13), respectively. The other terms in (36) have been 
calculated earlier. In fact, in terms of the symmetry transformations,
(36) can be recast as
$$
\begin{array}{lcl}
T_{\mu\nu}^{(s)} = \frac{i}{2} \;s_{b} \;\bar s_{b}\;
\bigl (A_{\mu} A_\nu - \frac{1}{2} \eta_{\mu\nu} A_\rho A^\rho \bigr )
- \frac{i}{2}\;s_{d}\;\bar s_{d}\;
\bigl (\varepsilon_{\mu\rho} \varepsilon_{\nu\sigma} 
A^{\rho}  A^\sigma + \frac{1}{2} \eta_{\mu\nu} A_\rho A^\rho \bigr ).
\end{array} \eqno(38)
$$
The geometrical interpretation for $T_{\mu\nu}^{(s)}$
in (36) can be provided in the same manner as the arguments and explanations
given for the Lagrangian density after equation (19b). It appears to be an
essential feature of a TFT that its symmetric energy momentum tensor can be
expressed as the $\theta\bar\theta$-component of a second-rank tensor that
can be constructed by the even superfields of the theory. On this component,
we apply the constraint conditions (8) and (13) that emerge after the 
imposition of the (dual) horizontality conditions.

It is gratifying to point out that, in the superfield formulation, the
symmetric form of the  energy momentum tensor, the expressions for $T_{(1,2)},
P_{(1,2)}$ in (17), the expressions for $S_{\mu\nu}^{(1,2)}$
and $\bar S_{\mu\nu}^{(1,2)}$, the correct form of the topological invariants,
etc., come out very naturally. Similarly,
the form of the Lagrangian density turns out to be the Grassmannian
derivatives on the Lorentz scalar ($B_\rho (x,\theta,\bar\theta)
B^\rho (x,\theta,\bar\theta)$) when we exploit the nilpotent (anti-)BRST- and
(anti-)co-BRST symmetries together with the generalized versions of
the horizontality condition. To be more precise
and more elaborate, it is the $\theta\bar\theta$-component of 
the above Lorentz scalar and the second rank tensors:
$B_\mu (x,\theta,\bar\theta) B_\nu (x,\theta,\bar\theta)$ 
and $\varepsilon_{\mu\rho} \varepsilon_{\nu\sigma}
B^\rho (x,\theta,\bar\theta) B^\sigma (x,\theta,\bar\theta)$, that leads to
the derivation of the Lagrangian density and the symmetric energy momentum
tensor. In this derivation, 
the generalized versions of horizontality condition w.r.t. the super
cohomological operators $\tilde d$ and $\tilde \delta$ play a very decisive
role.
Keeping in  mind the geometrical interpretations for the
(anti-)BRST charges $(\bar Q_{b})Q_{b}$ and (anti-)co-BRST charges
$(\bar Q_{d})Q_{d}$ as the translation generators,
it is obvious that the Lagrangian density in (17) (or its superfield
analogue (19a)) and the energy momentum tensor in (28) (or its superfield
analogue in (36)) can be thought of as the  translations of superfield
versions (cf.Eqns.(18a,18b)) of the local composite fields 
$T_{(1,2)}(P_{(1,2)})$  
and $S_{\mu\nu}^{(1,2)}(\bar S_{\mu\nu}^{(1,2)})$ along the Grassmannian 
directions of the $(2+2)$-dimensional supermanifold. These properties 
are some of the key requirements
for the existence of a TFT. Furthermore, it is also evident from
(26) and (31) that the zero-forms of the topological invariants and
$P's$ and $T's$ of (17)  are nothing but the Grassmannian
($\theta$ and $\bar \theta$) components in the expansion of the
superfields $\Phi \bar \Phi$. Geometrically, 
the zero-forms of the topological invariants are nothing but the translations
of the local (but composite) fields $(\Phi \bar\Phi) (x,\theta,\bar\theta)$
along the $\theta$- and $\bar\theta$ directions of the $(2+2)$-dimensional
supermanifold. It would be nice to apply this superfield
formalism to the case of 2D self-interacting non-Abelian gauge theory
and 4D free Abelian two-form gauge theory where the existence of nilpotent
(anti-)BRST- and (anti-)co-BRST symmetries have been demonstrated. Such
studies might turn out to be useful in the context of topological string
theories and topological gravity where, in contrast to the flat Minkowskian
metric of our present discussion, a non-trivial (spacetime-dependent)
metric is considered for the sake of generality. These
are some of the issues that are under investigation and our results will be
reported elsewhere [28].\\

\noindent
{\bf Acknowledgements}\\

\noindent
Some interesting and stimulating comments by the referee on the need to
obtain an Euclidean version of the 2D Minkowskian gauge theories on the 2D
closed Riemann surfaces, its subsequent topological and mathematical
implications, etc., are gratefully acknowledged.

\end{document}